\documentclass[graybox]{svmult}

\newcommand       \Angstrom     {\,{\rm \AA}}
\newcommand       \kpc          {\,{\rm kpc}}

\newcommand       \mum          {\,{\rm \mu m}}

\newcommand       \Ks           {{\rm K_{S}}}

\newcommand       \HH           {{\rm H}}

\newcommand       \simali       {\,{\sim}}
\newcommand       \magni        {\,{\rm mag}}

\newcommand       \RV           {{R_V}}
\newcommand       \AV           {{A_V}}
\newcommand       \Rv           {{R_V}}

\newcommand       \magkpc       {\,{\rm mag\, kpc}^{-1}}

\newcommand       \JJ           {{\rm J}}

\newcommand       \ppm          {\,{\rm ppm}}
\usepackage{mathptmx}       
\usepackage{helvet}         
\usepackage{courier}        
\usepackage{type1cm}        
\usepackage{makeidx}         
\usepackage{graphicx}        
\usepackage{multicol}        
\usepackage[bottom]{footmisc}

\makeindex             

\begin{document}

\title*{Dust in the Local Group}
\author{Aigen Li$^1$, 
             Shu Wang$^2$, 
             Jian Gao$^2$, 
    \and B.W.~Jiang$^2$}
\authorrunning{Li, Wang, Gao \& Jiang}
\institute{1.~Department of Physics and Astronomy,
           University of Missouri,
           Columbia, MO 65211, USA
\and 2.~Department of Astronomy, Beijing Normal University,
     Beijing 100875, China
}
%
%
\maketitle

\abstract{How dust absorbs and scatters starlight
as a function of wavelength (known as the interstellar
extinction curve) is crucial for correcting for
the effects of dust extinction in inferring
the true luminosity and colors of reddened astrophysical
objects. Together with the extinction spectral features,
the extinction curve contains important information
about the dust size distribution and composition.
This review summarizes our current knowledge of
the dust extinction of the Milky Way,
three Local Group galaxies
(i.e., the Small and Large Magellanic Clouds, and M\,31),
and galaxies beyond the Local Group.
}

\section{Introduction\label{sec:into}}
Interstellar dust is an important constituent of
the Milky Way (MW) and external galaxies
although it makes up only about 0.1\%
of the visible matter of a typical galaxy.
Understanding dust and its role in the universe 
are very important. Virtually all observations of 
astrophysical objects and their physical processes 
are affected by the presence of dust either within 
the system being studied or along its line of sight.   
Dust absorbs and scatters starlight efficiently in
the ultraviolet (UV), optical and near infrared (IR)
wavelength range. It is the dominant opacity source
for continuum photons with wavelengths longward of
the ionization edge of hydrogen.
In order to infer the intrinsic properties
(e.g., luminosity, color, spectral energy distribution)
of an astrophysical object,
it is crucial to correct for the effects of
dust extinction --- the sum of absorption and
scattering. Correcting for dust extinction is
also essential for inferring the stellar content
of a galaxy, or the history of star formation
in the universe.
The uncertainty in the correction for 
dust extinction currently dominates 
the uncertainty in the inferred
star formation rate in high-$z$ galaxies
(see Madau \& Dickinson 2014).

Dust absorbs and scatters starlight differently
at different wavelengths,
often with shorter-wavelength (blue) light
more heavily obscured than
longer-wavelength (red) light.
Therefore, dust controls the appearance of
a galaxy by dimming and reddening the starlight
in the UV and optical
(e.g., see Block \& Wainscoat 1991,
Block et al.\ 1994, Block 1996).
Dust re-radiates the absorbed UV/optical
stellar photons at longer wavelengths.
Nearly half of the bolometric luminosity
of the local universe is reprocessed by dust into
the mid- and far-IR (Dwek et al.\ 1998).

Dust is an important agent of
the galactic evolution.
Dust plays an important role in interstellar chemistry
by providing surfaces for the formation of the most
abundant molecule (i.e., H$_2$)
and by reducing the stellar UV radiation
which otherwise would photodissociate molecules.
The far-IR radiation of dust removes the gravitational
energy of collapsing clouds, allowing star formation
to occur. Dust dominates the heating of interstellar
gas through photoelectrons (Weingartner \& Draine 2001a).


In this review we will focus on dust extinction.
The study of interstellar extinction has a long
history (see Li 2005).
As early as 1847, Wilhelm Struve already
noticed that the apparent number of
stars per unit volume of space declines
in all directions receding from the Sun.
He attributed this effect to
interstellar absorption.
From his analysis of star counts
he deduced an visual extinction
of $\simali$1$\magkpc$.
In 1904, Jacobus Kapteyn
estimated the interstellar absorption
to be $\simali 1.6\magkpc$,
in order for the observed distribution
of stars in space to be consistent with
his assumption of a constant stellar density
(see van der Kruit 2014).
This value was amazingly close to the current estimates of
$\simali$1.8$\magkpc$ (see Li 2005).

How the interstellar extinction $A_\lambda$
varies with wavelength $\lambda$
--- known as the ``interstellar extinction law''
or ``interstellar extinction curve'' ---
is one of the primary sources of information
about the dust size distribution,
and the extinction spectral features provide
direct information on the dust composition (Draine 2003).
The interstellar extinction curve is most commonly
determined using the ``pair method'' which compares
the spectrum of a reddened star with that of
an unreddened star of the same spectral type
(see Appendix~1 in Li \& Mann 2012).
This method, first used by Bless \& Savage (1970),
has since been used extensively to deduce
the extinction curves for a large number of
Galactic sightlines (see Valencic et al.\ 2004
and references therein).
However, direct measurements of extragalactic
UV extinction using individual reddened stellar
sightlines are still limited to three Local Group
galaxies: the Small Magellanic Cloud (SMC),
the Large Magellanic Cloud, and M\,31
(see Clayton 2004).
In galaxies beyond the Local Group,
individual stars can not be observed.
Various approaches other than the ``pair method''
have been taken to determine the extinction curves.

In \S2 we summarize our current knowledge of
the UV, optical and IR extinction properties
of the MW.
The extinction curves of the SMC, LMC,
M\,31 are respectively discussed in
\S3, \S4, and \S5.
We summarize in \S6
the approaches to determining the extinction curve
for dust in galaxies where individual stars
cannot be resolved.
We will not discuss the dust thermal IR/mm emission.
The importance of the dust thermal emission 
of the MW (see Draine \& Li 2001, 
Li \& Draine 2001, Li 2004, Jones et al.\ 2013),
the SMC/LMC (Li \& Draine 2002; Galliano et al.\ 2011;
Gordon et al.\ 2011, 2014; Meixner et al.\ 2010, 2013),
and other galaxies (e.g., see Draine \& Li 2007, 
Draine et al.\ 2007) cannot be over-emphasized.
A simultaneous investigation of 
the interstellar extinction, scattering, polarization 
and thermal emission would place powerful constraints 
on the nature of the dust
(e.g., see Li \& Greenberg 1997, 1998; 
Block et al.\ 1997; Block \& Sauvage 2000;
Hensley \& Draine 2014).

\begin{figure}
\centering
\includegraphics[height=8cm,width=8cm]{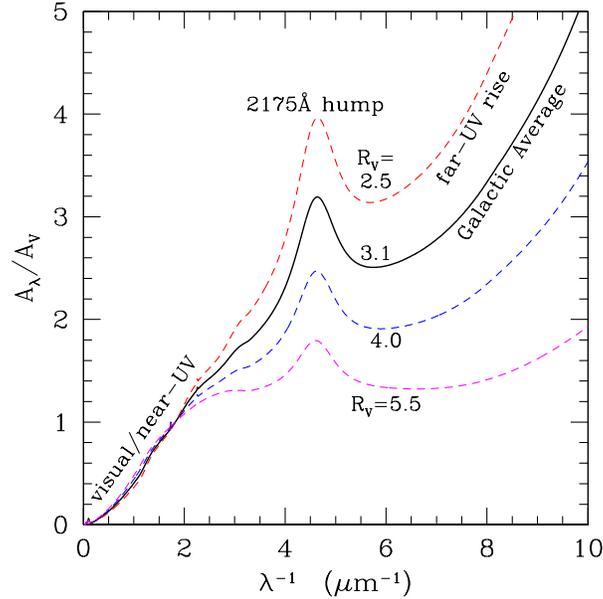}
\caption{\footnotesize
        Interstellar extinction curves of
        the Milky Way ($R_V$\,=\,2.5, 3.1, 4.0, 5.5).
        There exist considerable regional variations
        in the Galactic optical UV extinction curves,
        as characterized by 
        the total-to-selective
        extinction ratio $R_V$, indicating that
        dust grains on different sightlines have
        different size distributions.
        }
\end{figure}


\section{The Milky Way}
\subsection{UV/Optical Extinction:
            Strong Dependence on Environments
            }
The interstellar extinction curve
is usually expressed as $A_\lambda/\AV$,
where $\AV$ is the extinction in the $V$ band.
This means of expressing the extinction curve
is not unique; it has also been a common practice
to use instead the ratios of two colors,
$E(\lambda-V)/E(B-V)$,
where $E(\lambda-V)\equiv A_\lambda - \AV$,
and $A_B$ is the extinction in the blue ($B$) band.
The Galactic interstellar extinction curves
have now been measured for a large number of
sightlines over a wide wavelength range
(0.1$\mum$\,$<$\,$\lambda$\,$<$20$\mum$).
As shown in Figure~1, the extinction curves plotted
against the inverse wavelength $\lambda^{-1}$
rise almost linearly from the near-IR to the near-UV,
with a broad absorption bump at about
$\lambda^{-1}$$\approx$4.6$\mum^{-1}$
($\lambda$$\approx$2175$\Angstrom$)
and followed by a steep rise into the far-UV
at $\lambda^{-1}$$\approx$10$\mum^{-1}$,
the shortest wavelength at which the extinction
has been measured. 
Note the extinction curve does not
show any sign of a turnover in the far-UV rise.

The Galactic extinction curves are known 
to vary significantly among sightlines
in the UV/optical at $\lambda<0.7\mum$, 
including strong and weak 2175$\Angstrom$ bumps, 
and steep and flat far-UV extinction (see Figure~1).
Cardelli et al.\ (1989) found that the variation
in the UV/optical can be characterized
as a one-parameter function (hereafter ``CCM'') of
$\RV\,\equiv\,\AV/E(B-V)$,
the optical total-to-selective extinction ratio.
The value of $\RV$ depends upon the environment
along the line of sight. Low-density regions usually
have a rather low value of $\RV$,
having a strong 2175$\Angstrom$ bump
and a steep far-UV rise
at $\lambda^{-1}$\,$>$\,4$\mum^{-1}$.
Lines of sight penetrating into dense clouds,
such as the Ophiuchus or Taurus molecular clouds,
usually have $4 < \RV < 6$,
showing a weak 2175$\Angstrom$ bump,
and a relatively flat far-UV rise.
The ``average" extinction law for
the Galactic diffuse interstellar medium (ISM)
is described by a CCM extinction curve
with $\RV\approx 3.1$,
which is commonly used to correct observations
for dust extinction.
Theoretically, $\RV$ may become infinity in
dense regions rich in very large, ``gray'' grains
(i.e., the extinction caused by these grains does
not vary much with wavelength), while the steep
extinction produced completely by Rayleigh scattering
would have $\RV\sim 0.72$ (Draine 2003).
A wide range of $\RV$ values have been reported
for extragalactic lines of sight,
ranging from $\RV\approx 0.7$ for
a quasar intervening system
at $z\approx1.4$ (Wang et al.\ 2004)
to $\RV\approx 7$ for gravitational
lensing galaxies (Falco et al.\ 1999).
However, the one-parameter CCM formula of $\RV$
does not apply to sightlines beyond the MW
(see Clayton 2004).

The exact nature of the carrier of
the 2175$\Angstrom$ extinction bump
remains unknown since its first discovery
nearly half a century ago (Stecher 1965).
Recently, a popular hypothesis is that it
is due to the $\pi$\,--$\pi^{\ast}$ transition
of polycyclic aromatic hydrocarbon (PAH) molecules
(Joblin et al.\ 1992, Li \& Draine 2001,
Cecchi-Pestellini et al.\ 2008, Malloci et al.\ 2008,
Steglich et al.\ 2010, Mulas et al.\ 2013).

\begin{figure}[h!]
\centering
\includegraphics[angle=0,width=12.5cm]{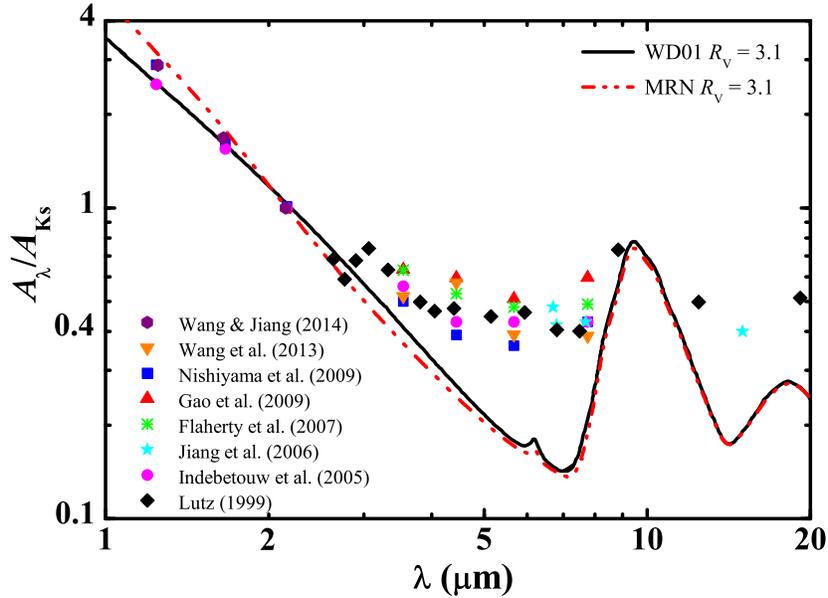}
\caption{\footnotesize
               Comparison of the IR extinction observed for various
               interstellar regions with that predicted from the MRN
               (dot-dashed line) and WD01 (solid line)
               silicate-graphite models for the diffuse ISM of which
               the UV/optical extinction is characterized by
               $R_V\approx 3.1$.
               The little bump at 6.2$\mum$ arises from
               the C--C stretching absorption band of PAHs
               (see Li \& Draine 2001).
               }
\end{figure}

\subsection{Near-IR Extinction: A Universal Power Law?}
With the wealth of available data from space-borne telescopes
(e.g., {\it ISO} and {\it Spitzer}) 
and ground-based surveys (e.g., {\it 2MASS})
in the near- and mid-IR, 
in recent years we have seen an increase
in interest in IR extinction. 
Understanding the effects of
dust extinction in the IR wavelengths 
is important to properly interpret
these observations.

As the UV/optical extinction varies substantially 
among various environments, 
one might expect the near-IR extinction
at $0.9\mum < \lambda <3\mum$ to vary correspondingly.
However, for over two decades astronomers 
have believed that there is little, 
if any, near-IR extinction variation
from one line of sight to another.
The near-IR extinction law appears to be an 
approximately uniform power law of 
$A_\lambda \sim \lambda^{-\alpha}$
with $\alpha$\,$\approx$\,1.6--1.8,
independent of environment or $R_{\rm V}$
at $0.9\mum < \lambda < 3\mum$.
   Martin \& Whittet (1990) found $\alpha\approx1.8$
   in the diffuse ISM 
   as well as the outer regions of
   the $\rho$ Oph and Tr\ 14/16 clouds.
   Using a large sample of obscured OB stars,
   He et al.\ (1995) found that the near-IR extinction
   can be well-fitted to a power law
   with $\alpha\approx 1.73\pm0.04$,
   even though $R_{\rm V}$ varies between 2.6 and 4.6.
A ``universal'' power law also well describes
the near-IR polarization $P_\lambda$ which,
like the near-IR extinction,
probes the large, submicron-sized grain population,
$P_\lambda/P_{\rm max}\propto \lambda^{-1.8}$,
where $P_{\rm max}$ is the peak polarization
(Martin \& Whittet 1990).

However, much steeper power-laws
have also been derived
for the near-IR extinction.
Stead \& Hoare (2009) determined
$\alpha\approx 2.14^{+0.04}_{-0.05}$
for the slope of the near-IR extinction power law
for eight regions of the MW
between $l \sim 27^{\rm o}$ and $\sim 100^{\rm o}$.
Nishiyama et al.\ (2009) explored
the extinction law toward the Galactic center.
They derived the index of the power law $\alpha\approx1.99$.
Fritz et al.\ (2011) found $\alpha\approx2.11$
for the Galactic center extinction.

Based on  a spectroscopic study of
the 1--2.2$\mum$ extinction law
toward a set of nine ultracompact HII regions
with $\AV>15\magni$, Moore et al.\ (2005) found
some evidence that the near-IR extinction
curve may tend to flatten at higher extinction.
They argued that flatter curves are most likely
the result of increasing the upper limit of
the grain-size distribution in regions of
higher extinction.
Naoi et al.\ (2006) determined the near-IR color excess ratio
$E(\JJ-\HH)/E(\HH-\Ks)$, one of the simplest parameters for
expressing the near-IR extinction law, for L1688, L1689,
and L1712 in  the $\rho$ Oph cloud, and Cha I, Cha II, and Cha III
in the Chamaeleon cloud. They found that $E(\JJ-\HH)/E(\HH-\Ks)$
changes with increasing optical depth,
consistent with grain growth
toward the inside of the dark clouds.

More recently, Wang \& Jiang (2014) examined the apparent
colors of 5,942 K-type giants whose intrinsic J, H, $\Ks$
colors are known from their effective stellar temperatures
determined from the SDSS-III APOGEE survey.
They derived a nearly constant near-IR color excess
ratio of $E(\JJ-\HH)/E(\HH-\Ks)\approx0.64$
(which corresponds to $\alpha\approx 1.95$),
independent of the amount of extinction
in the color excess range of $0.3 < E(\JJ-\Ks) < 4.0$.

A ``universal'' near-IR extinction law of
a constant power-index $\alpha$ for regions 
with different physical conditions is difficult
to understand. One would imagine that in denser
regions the dust could be larger which would lead to
a smaller $\alpha$. However, this tendency is not
observationally verified 
(see Figure~7 of Wang et al.\ 2013).
 %

\begin{figure}[h!]
\hspace{-0.4in}
\includegraphics[angle=0,width=14.0cm]{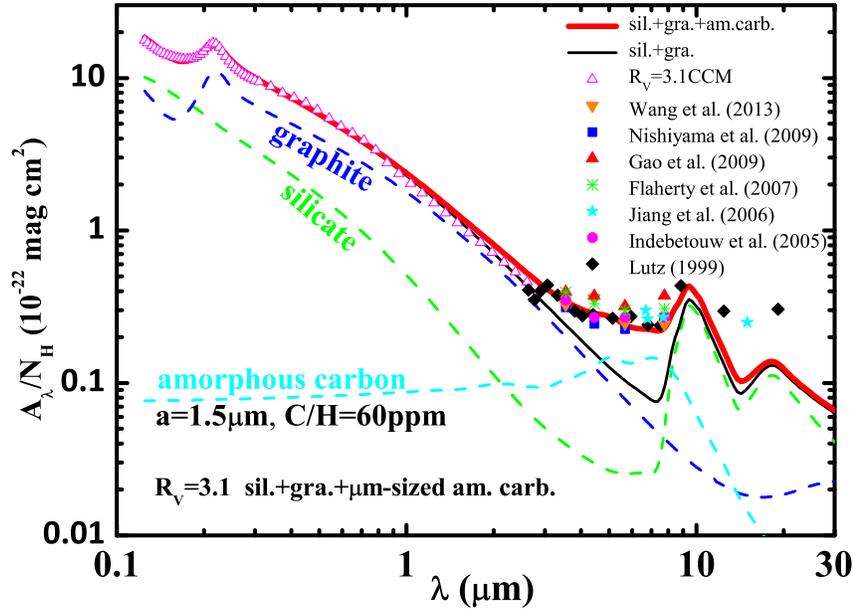}
\caption{\footnotesize
               Fitting the $R_V=3.1$ UV/optical,
               near- and mid-IR extinction
               with a simple mixture (dot-dashed line)
               of amorphous silicate (dotted line)
               and graphite dust (dashed line),
               together with a population of
               large, $\mu$m-sized
               amorphous carbon of
               spherical radii $a\approx 1.5\mum$
               and C/H\,$\approx$\,60$\ppm$
               (Wang, Li \& Jiang 2014).
               The thick solid line plots the model fit,
               while the symbols plot the observed extinction:
               the open pentagons plot the $\RV=3.1$
               UV/optical/near-IR extinction, 
               and the other symbols
               plot the mid-IR extinction.
             }
\end{figure}

\subsection{Mid-IR Extinction: Universally Flat?}
The mid-IR extinction at $3\mum <\lambda<8\mum$
(in between the power-law regime
at $\simali$1--3$\mum$ and the 9.7$\mum$
silicate Si--O stretching absorption feature)
is not well understood.
The determination of the mid-IR extinction law
has been more difficult because this wavelength
range is best accessed from space.
All dust models for the diffuse ISM predict
an extinction curve steeply declines with $\lambda$
at $1\mum < \lambda < 7\mum$
(at $\lambda>7\mum$, the extinction increases
because of the 9.7$\mum$ silicate absorption band).
As shown in Figure~2,
the Weingartner \& Draine (2001b; WD01)
silicate-graphite grain model predicts a power-law of
$A_\lambda \propto \lambda^{-1.74}$ for the IR extinction
at $1\mum < \lambda < 7\mum$,
while the Mathis, Rumpl, \& Nordsieck (1977; MRN)
model predicts a steeper power-law of
$A_\lambda \propto \lambda^{-2.02}$.
The model IR extinction curves reach their minimum at
$\simali$7$\mum$ where the extinction power-law intersects
the blue-wing of the 9.7$\mum$ silicate absorption band.

Rieke \& Lebofsky (1985) measured the IR extinction
from 1$\mum$ to 13$\mum$ for the lines of sight
toward $o$  Sco, a normal A5\,II star behind the edge
of the $\rho$ Oph cloud obscured by
$A_V\approx 2.92\magni$,
and toward a number of stars in the Galactic center.
They derived a power-law of
$A_\lambda\propto \lambda^{-1.62}$ for
$1\mum < \lambda < 7\mum$ for
$o$ Sco and the Galactic center sources.
Draine (1989) compiled the IR extinction observed
for a range of Galactic regions 
including diffuse clouds,
molecular clouds, and HII regions.
A power-law of $A_\lambda\propto \lambda^{-1.75}$ for
$1\mum < \lambda < 7\mum$ was obtained.
Bertoldi et al.\ (1999) and Rosenthal et al.\ (2000)
also derived a power-law extinction of
$A_\lambda\propto \lambda^{-1.7}$
for $2\mum <\lambda < 7\mum$
for the Orion molecular cloud (OMC)
which displays an absorption band at 3.05$\mum$
attributed to water ice.

However, numerous recent observations suggest that
the mid-IR extinction at $3\mum <\lambda< 8\mum$
appears to be almost {\it universally} flat or ``gray''
for both diffuse and dense environments,
much flatter than that predicted from
the MRN or WD01 silicate-graphite model
for $R_V=3.1$ (see Figure~2).

Lutz et al.\ (1996) derived the extinction 
toward the Galactic center star Sgr A$^{\ast}$ 
between 2.5$\mum$ and 9$\mum$
from the H recombination lines. 
They found that the Galatic center extinction 
shows a flattening of $A_\lambda$ 
in the wavelength region of 
$3\mum <\lambda < 9\mum$,
clearly lacking the pronounced dip at $\simali$7$\mum$
predicted from the $R_V=3.1$ silicate-graphite model
(see Figure~2). This was later confirmed by Lutz (1999),
Nishiyama et al.\ (2009), and Fritz et al.\ (2011).

Indebetouw et al.\ (2005) used the photometric data
from the {\it 2MASS} survey and the {\it Spitzer}/GLIMPSE
Legacy program to determine the IR extinction.
From the color excesses of background stars,
they derived  the $\simali$1.25--8$\mum$ extinction laws
for two very different lines of sight in the Galactic plane:
the $l=42^{\rm o}$ sightline toward a relatively quiescent region,
and the $l=284^{\rm o}$ sightline which crosses the Carina Arm
and contains RCW~49, a massive star-forming region.
The extinction laws derived for 
these two distinct Galactic plane
fields are remarkably similar:
both show a flattening across 
the 3--8$\mum$ wavelength range,
consistent with that derived by 
Lutz et al.\ (1996) for the Galactic center.
%
%

Jiang et al.\ (2006) derived the extinction at 7 and 15$\mum$
for more than 120 sightlines in the inner Galactic plane based
on the ISOGAL survey data and the near-IR data from DENIS and 2MASS,
using RGB tip stars or early AGB stars (which have only moderate mass
loss) as the extinction tracers. They found the extinction  well
exceeding that predicted from the MRN or WD01 $R_V=3.1$ models.

Flaherty et al.\ (2007) obtained the mid-IR
extinction laws in the {\it Spitzer}/IRAC bands
for five nearby star-forming regions.
The derived extinction laws
at $\simali$4--8$\mum$ are flat, even flatter than
that of Indebetouw et al.\ (2005).

Gao, Jiang, \& Li (2009) used the {\it 2MASS}
and {\it Spitzer}/GLIPMSE data to derive
the extinction in the four IRAC bands
for 131 GLIPMSE fields
along the Galactic plane within $|l|\leq65^{\rm o}$
(Benjamin et al.\ 2003, Churchwell et al.\ 2006).
Using red giants and red clump giants 
as the extinction tracers,
they also found the mean extinction
in the IRAC bands to be flat.

Wang et al.\ (2013) determined the mid-IR extinction
in the four {\it Spitzer}/IRAC bands of five individual
regions in Coalsack, a nearby starless dark cloud,
spanning a wide variety of interstellar environments
from diffuse and translucent to dense clouds.
They found that all regions exhibit a flat mid-IR extinction.

All these observations appear to suggest
an ``universally'' flat extinction law in the mid-IR,
with little dependence on environments.
Draine (2003) extended the WD01 model of $\RV=5.5$
into the IR up to $\lambda < 30\mum$ and found that
the WD01 $\RV=5.5$ model closely reproduces
the flat mid-IR extinction observed toward
the Galactic center (Lutz et al.\ 1996).
Dwek (2004) hypothesized that the observed
mid-IR extinction is largely due to metallic
needles for which the opacity is high
in the mid-IR.
Wang, Li, \& Jiang (2014) found that the mid-IR
extinction as well as the UV/optical extinction could
be closely reproduced by a mixture of
submicron-sized amorphous silicate and graphite,
and a population of micrometer-sized amorphous carbon
which consumes $\simali$60 carbon atoms
per million (ppm) hydrogen atoms (see Figure~3).

\begin{figure}[ht]
  \begin{center}
  \includegraphics[angle=0,width=12.5cm]{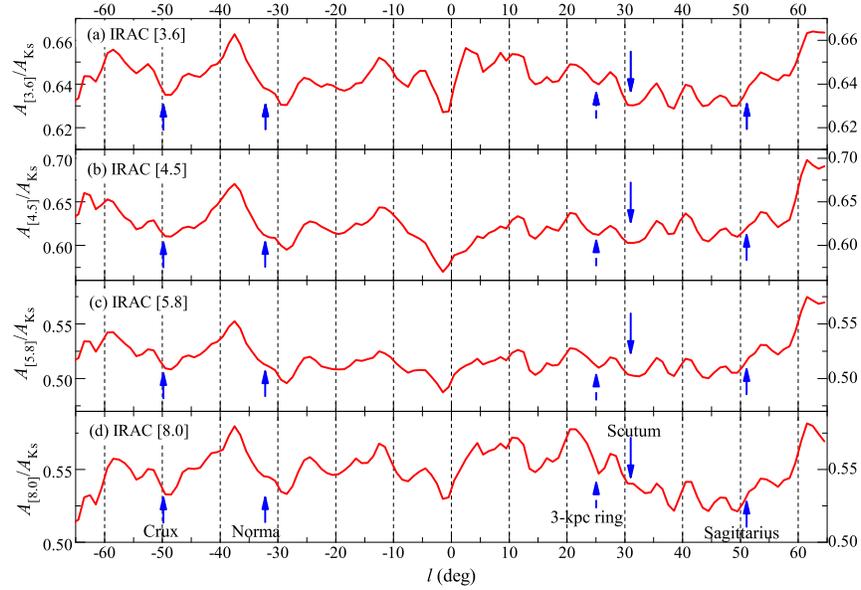}
  \end{center}
\caption{\footnotesize
         Longitudinal distributions of $A_{\lambda}/A_\Ks$ in
         the four IRAC bands 
         (solid lines; Gao, Jiang \& Li 2009).  
         The dips of the extinction ratios
         $A_\lambda/A_\Ks$ appear to coincide
         with the locations
         of the spiral arms.
         The solid vertical arrows
         show the tangent directions of the spiral arms at
         $l=-50^{\circ}, -33^{\circ}, 31^{\circ}$
         and $51^{\circ}$ (Vall\'ee 2008).
         The dashed arrow shows the tangent direction of
         the 3$\kpc$ ring at $l=23^{\circ}$
         (Dame \& Thaddeus 2008).
	 }
\vspace{-3mm}
\end{figure}

%
\subsection{IR Extinction as a Probe of
            the Galactic Structure}
Whittet (1977) presented observational evidence
for a small but appreciable variation in $\RV$
with Galactic longitude. He suggested that the most
likely explanation for this is a variation
in the mean size of the dust in the local spiral arm.
However, unfortunately only a few data points were used
in that work and therefore no systematic variation of
the extinction with Galactic longitude was reported.
A major leap forward occurred recently thanks to
the amazing success of the {\it Spitzer}/GLIMPSE 
Legacy programs (see Benjamin 2014).

\begin{figure}[ht]
  \begin{center}
  \includegraphics[width=8.0cm]{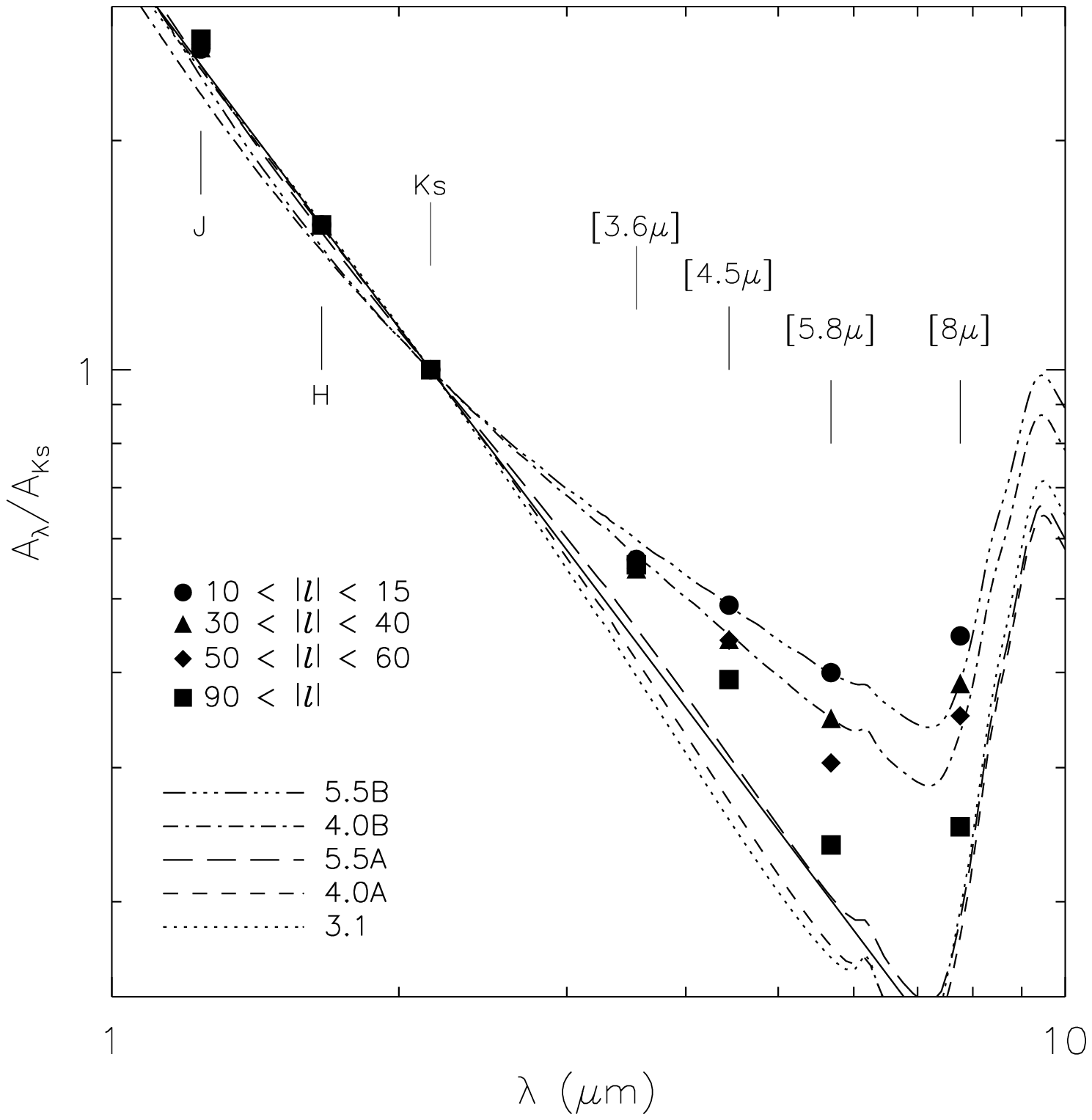}
  \end{center}
\caption{\footnotesize
         Mean extinction curves for the indicated ranges
         of Galactic angle
         (circles: $10^\circ < |l| < 15^\circ$;
          triangles: $30^\circ < |l| < 40^\circ$;
          diamonds: $50^\circ < |l| < 60^\circ$;
          squares: $|l| > 90^\circ$),
          compared to that of the WD01 model for
          $R_V$\,=\,3.1, 4.0, 5.5.
          Also shown for comparison is an
          $A_\lambda\propto\lambda^{-\beta}$ power law,
          with $\beta$\,=\,1.66 (solid line).
          It is apparent that the extinction curve
          becomes increasingly steep at larger
          Galactocentric angles
          (Zasowski et al.\ 2009).
          }
\vspace{-2mm}
\end{figure}

Based on the IR extinction in the four IRAC bands
derived for 131 GLIPMSE fields
in the Galactic plane,
Gao, Jiang, \& Li (2009) demonstrated that
the mid-IR extinction appears to vary with
Galactic longitude; more specifically,
the locations of the spiral arms seem to
coincide with the dips of the extinction ratios
$A_\lambda/A_\Ks$ (see Figure~4).
In particular, the coincidence of the locations of
the $A_\lambda/A_\Ks$ minimum and the spiral arms is
outstanding at negative longitudes
(e.g., the Crux-Scutum arm at $l=-50^{\circ}$,
the Norma arm at $-33^{\circ}$,
and the southern tangent of the 3$\kpc$ ring
around $-23^{\circ}$). For the positive longitudes,
the locations of the Scutum-Crux arm
and the Sagittarius-Carina arm also
coincide with the valleys of $A_\lambda/A_\Ks$.
For the broad dip around $l=16^{\circ}$,
there is no clear tangent
direction to any arms,
but this direction points to the start of
the Norma arm and the Scutum-Crux arm
(see Figure~2 of Vall\'ee 2008),
and probably one end of the Galactic bar.

Using data from \emph{2MASS} and \emph{Spitzer}/IRAC
for G and K spectral type red clump giants,
Zasowski et al.\ (2009) also found longitudinal variations
of the 1.2--8$\mum$ IR extinction over $\simali$150$^{\rm o}$
of contiguous Galactic mid-plane longitude;
more specifically, they found strong, monotonic
variations in the extinction law shape as a function
of angle from the Galactic center, symmetric on either
side of it: the IR extinction law becomes increasingly
steep as the Galactocentric angle increases,
with identical behavior between $l < 180^{\rm o}$
and  $l > 180^{\rm o}$ (see Figure~5).

\begin{figure}
\centering
\includegraphics[angle=0,width=5in]{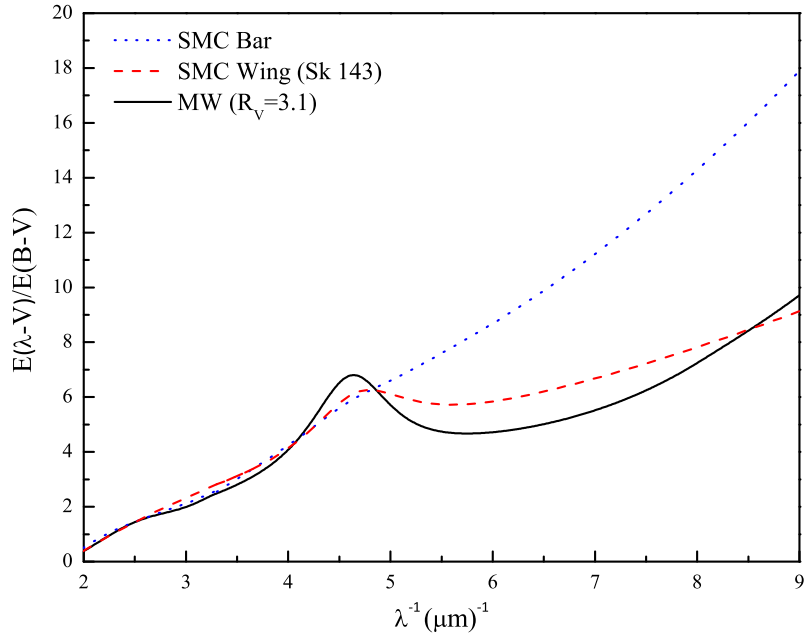}
\caption{\footnotesize
         Comparison of the SMC Bar (dotted)
         and MW $\RV=3.1$ (solid)
         extinction curves.
         Also shown is the extinction curve for 
         the sightline toward the SMC wing star
         Sk 143 (dashed).
         }
\end{figure}

\section{SMC}
As our nearest galactic neighbors, 
the Magellanic Clouds offer a unique opportunity 
to study the effects of different galactic environments 
on dust properties. 
As a metal-poor (with a metallicity only $\sim$1/10 of
that in the MW; see Kurt \& Dufour 1998)
and gas-rich (with a dust-to-gas ratio over 10 times
lower than in the MW; see Bouchet et al.\ 1985)
irregular dwarf galaxy, the SMC is often considered as
a local analog of primordial galaxies
at high redshifts which must have formed at very low metallicity.
Therefore, the dust in the SMC which differs substantially from
that in the MW allows us to probe
the primordial conditions in more distant galaxies.
Indeed, the SMC-type extinction often provides 
better fits to the spectral energy distributions 
(SEDs) of many extragalactic systems 
than the ``standard'' MW-type extinction curve
(e.g., see Richards et al.\ 2003 and Hopkins et al.\ 2004
for reddened quasars, and Vijh et al.\ 2003 for 
Lyman break galaxies). The 2175$\Angstrom$ feature
seen in the MW-type extinction is absent in 
the attenuation curve inferred from radiative 
transfer modeling of starburst galaxies
including stars, gas and dust 
(Calzetti et al.\ 1994, Gordon et al.\ 1997).

Due to its low metallicity, the dust quantity (relative to H)
in the SMC is expected to be lower than that of the MW because
there is less raw material (i.e., heavy elements) available for
making the dust. The (relative) lack of the dust-making raw material
could prevent the dust in the SMC from growing and hence the dust
in the SMC may be smaller than the MW dust.
Furthermore, the star-formation activity
in the SMC could destroy the dust.
Therefore, one would naturally expect the dust size distribution
and extinction curve in the SMC to differ from that of the MW.

There are significant regional variations
in the UV/optical extinction properties of
the SMC. As shown in Figure~6,
the extinction curve of the SMC Bar
displays a nearly linear rise with
$\lambda^{-1}$
and no detectable 2175$\Angstrom$ extinction bump
(
Lequeux et al.\ 1982;
Cartledge et al.\ 2005),
presumably due to the destruction of the carriers
of the 2175\AA\ hump by the intense UV
radiation and shocks associated with star formation.
In contrast, the extinction curve for
the line of sight toward Sk 143 (AvZ 456)
has a strong 2175\AA\ hump (Lequeux et al.\ 1982;
  Cartledge et al.\ 2005).
This sightline passes through the SMC wing,
a region with much weaker star formation
(Gordon \& Clayton 1998).

The overall IR emission spectrum of the SMC
peaks at $\lambda$$\simali$100$\mum$
with a local minimum at $\lambda$$\simali$12$\mum$
which is commonly believed to be emitted by PAHs
(see Li \& Draine 2002).
%
The PAH emission features have been detected
locally in the SMC B1\#1 quiescent molecular cloud
(Reach et al.\ 2000),
and in some star-forming regions
(Contursi et al.\ 2000, Sanstrom et al.\ 2012).
Based on the observed PAH emission
and attributing the 2175$\Angstrom$
extinction bump to PAHs,
Li \& Draine (2002) predicted an extinction
excess of $\Delta A_{2175}\approx0.5\magni$
at 2175$\Angstrom$ for sightlines through SMC B1\#1.
Ma{\'{\i}}z Apell{\'a}niz \& Rubio (2012)
measured the UV extinction for four stars
in the SMC B1\#1 cloud from their HST/STIS
slitless UV spectra. They found that the 2175$\Angstrom$
bump is moderately strong in one star, weak in two stars,
and absent in the fourth star.

\begin{figure}
\centering
\includegraphics[angle=0,width=5in]{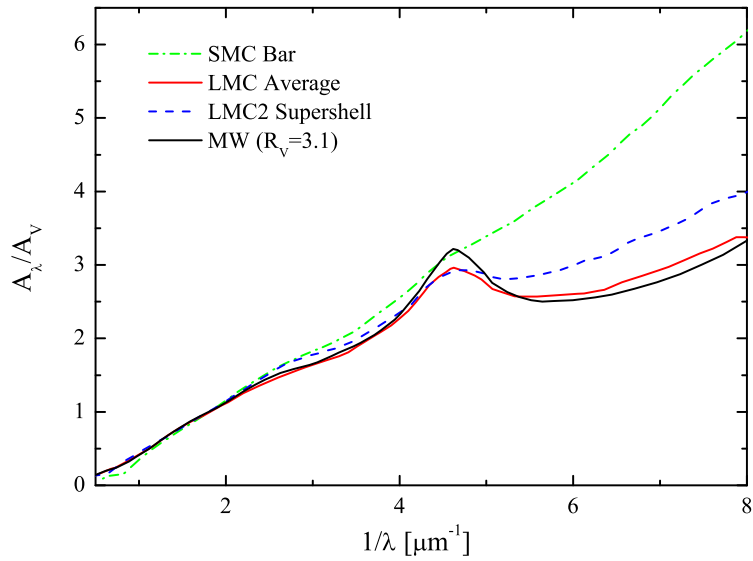}
\caption{\footnotesize
         Comparison of the ``LMC Average'' (dashed),
         SMC (dotted), and MW $\RV=3.1$ (solid)
         extinction curves.
         Also shown is the LMC2 supershell
         extinction curve (dot-dashed).
         The ``LMC Average'' curve 
         (also known as the ``LMC'' curve) 
         is the mean extinction curve 
         derived for the sightlines toward 
         the stars outside of the 30 Dor 
         star-forming region, 
         while the LMC2 supershell curve
         (also known as the ``LMC2'' curve
          or the ``LMC 30 Dor'' curve) 
         is for the dust inside 
         the LMC2 supergiant shell which,
         lying on the southeast side of 30 Dor,
         was formed by the combined stellar winds 
         and supernova explosions 
         from the stellar association at its center.
         }
\end{figure}

\section{LMC}
Like the SMC, the LMC is also a low-metalicity
irregular dwarf galaxy and a satellite of the MW.
Since the metallicity of the LMC
(which is only $\simali$1/4 of that of the MW;
Russell \& Dopita 1992) is similar to that of
galaxies at red shifts $z\sim1$ (see Dobashi et al.\ 2008),
it offers opportunities to study the dust properties
in distant low-metallicity extragalactic environments
by studying the extinction properties of the LMC.

As illustrated in Figure~7,
the LMC extinction curve is intermediate between
that of the MW and that of the SMC:
compared to the Galactic extinction curve,
the LMC extinction curve is characterized by
a {\sl weaker} 2175$\Angstrom$ hump
and a {\sl stronger} far-UV rise
(Nandy 1981, Koornneef \& Code 1981,
Gordon et al.\ 2003).
Strong regional variations in extinction properties
have also been found in the LMC
(Clayton \& Martin 1985;
Fitzpatrick 1985, 1986;
Misselt et al.\ 1999;
Gordon et al.\ 2003):
the sightlines toward the stars inside or near
the supergiant shell, LMC\,2, which lies on
the southeast side of the 30 Doradus star-forming region,
have a weak 2175$\Angstrom$ hump
(Misselt et al.\ 1999),
while the extinction curves
for the sightlines toward the stars
which are $>$\,500\,pc away from
the 30 Doradus region are
closer to the Galactic extinction curve.

Based on the photometric data
from the \emph{Spitzer}/SAGE survey
and with red giants as the extinction tracer,
Gao et al.\ (2013) derived the mid-IR
extinction for a number of regions of
the LMC in the four IRAC bands.
%
The average mid-IR extinction shows a flat curve,
close to the MW $\RV=5.5$ model extinction curve
(see Figure~8).

\begin{figure}
\centering
\includegraphics[angle=0,width=5in]{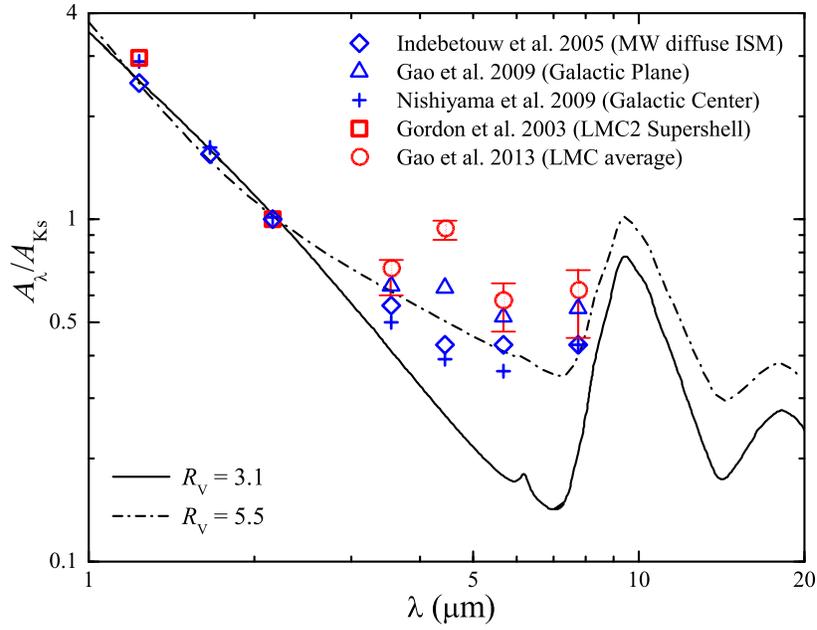}
\caption{\footnotesize
          IR extinction of the LMC
          compared with the WD01 model curves
          of $\Rv = 3.1$ (solid line)
          and $\Rv = 5.5$ (dot-dashed line).
          Open squares:
              the LMC near-IR extinction 
              of Gordon et al.\ (2003).
          Filled stars: 
              the mid-IR extinction at the four IRAC bands
              derived by Gao et al.\ (2013).
          Open diamonds: 
              the near- and mid-IR extinction
              for the MW diffuse ISM
              of Indebetouw et al.\ (2005).
          Open triangles: 
              the average extinction
              at the four IRAC bands
              derived from 131 GLIMPSE fields
              along the Galactic plane
              (Gao et al.\ 2009).
          Pluses: the near- and mid-IR extinction
                  toward the Galactic center
                  of Nishiyama et al.\ (2009).
          }
\end{figure}


In the 30 Dor star-forming region
where large amounts of UV radiation
and shocks are present, the dust is
expected to be processed and its size
distribution is expected to be modified.
The difference between the UV extinction 
characteristic of the 30 Dor region 
and that outside the 30 Dor region
could be understood in terms of dust
erosion in the 30 Dor region 
which produces a dust size distribution 
skewed toward smaller grains,
and leads to a steeper far-UV extinction
and a weaker 2175$\Angstrom$ bump 
as the bump carriers could also be destroyed. 

However, it is difficult to quantify 
the effects of star-formation activities
on the dust size distribution 
and the corresponding UV extinction.
Although the environment of the SMC Bar is likely 
to be less severe than for the 30 Dor region
(which is a much larger star-forming region 
than any in the SMC), the UV extinction of
the SMC Bar is more extreme than that of 
the LMC 30 Dor, with the former exhibiting
a very steep far-UV rise 
and lacking the 2175$\Angstrom$ bump.
The SMC has star formation occurring 
at only $\simali$1\% of the rate of 
a starburst galaxy. But starburst galaxies
appear to be dominated by large, submicron-
or micron-sized dust and are characterized 
by a flat or ``gray'' extinction curve
(see Calzetti 2001).
Therefore, many factors including 
densities, metallicities, 
and star formation rates
must be important in controlling
the formation and destruction processes 
(such as shattering and coagulation) of
dust (see Clayton 2004).

\begin{figure}
\centering
\includegraphics[angle=0,width=5in]{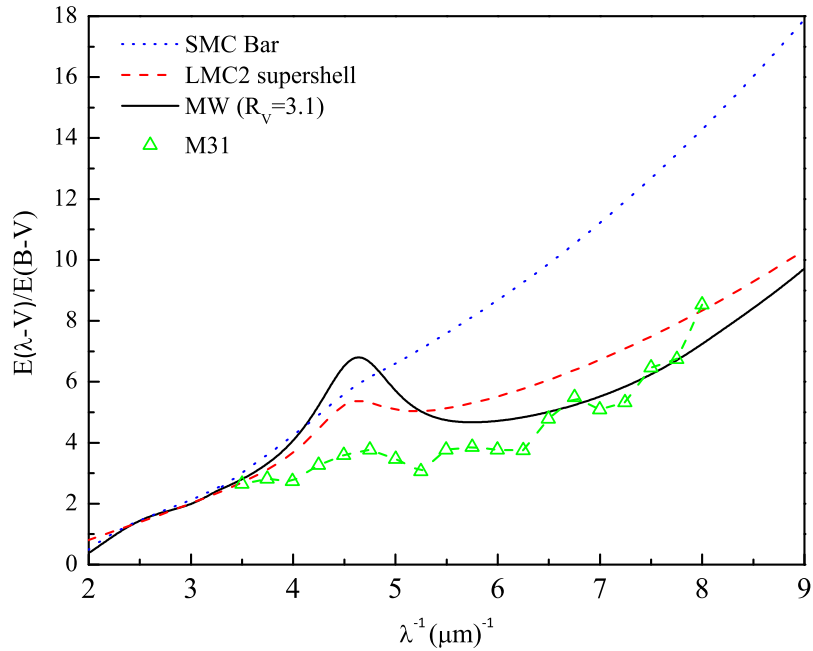}
\caption{\footnotesize
         Comparison of the M\,31 
         (dot-dashed line-connected triangles),
         MW $\RV=3.1$ (solid),
         SMC Bar (dotted), and 
         LMC2 30 Dor ``supershell' region (dashed)
         extinction curves.
         The M\,31 extinction curve was obtained
         by Bianchi et al.\ (1996) with
         the ``pair method'' by comparing the HST/FOS
         spectra of reddened M\,31 stars and 
         that of unredded Galactic stars
         of the same spectral type.
         The M\,31 extinction curve
         has an overall wavelength dependence 
         similar to that of the average Galactic 
         extinction curve 
         but possibly has a weaker 2175$\Angstrom$ bump. 
         }
\end{figure}

\section{M\,31}
%
%
The Andromeda galaxy (M\,31), 
at a distance of $\simali$780\,kpc
(McConnachie et al.\ 2005), 
is the only galaxy (other than 
the MW, and the SMC/LMC)
of which individual stars can 
be resolved and adopted for extinction
determination in terms of the ``pair method''.
Unlike the Magellanic Clouds which are metal-poor,
M\,31 has a super-solar metallicity.
%

%


Bianchi et al.\ (1996) obtained the UV spectra in 
the $\simali$1150--3300$\Angstrom$ wavelength range 
of several bright OB stars in 
M\,31, using the {\it Faint Object Spectrograph} 
(FOS) on board the {\it Hubble Space Telescope} (HST).
The UV extinction was derived for three sightlines
in M\,31 and seems to show a MW-type extinction curve 
but with the 2175$\Angstrom$ feature 
possibly somewhat weaker than in the MW
(see Figure~9).


More recently, Dong et al.\ (2014)
measured the extinction curve for the dust 
in the central $\simali$1$^{\prime}$
($\simali$200\,pc) region of M\,31 
at thirteen bands from the mid-UV 
to near-IR in the wavelength range
of $\simali$1928$\Angstrom$--1.5$\mum$.
They examined five representative dusty clumps
located in the circumnuclear region, 
utilizing data from the HST 
{\it Wide Field Camera 3} (WFC3) and 
{\it Advanced Camera for Surveys} (ACS) detectors
as well as the {\it UV and Optical Telescope} (UVOT)
on board {\it Swift}.
They found that the extinction curves
of these clumps, 
with $\RV\approx$\,2.4--2.5, 
are steeper than the average Galactic one,
indicating that the dust in M\,31 
is smaller than that in the MW.
Dong et al.\ (2014) also found that
one dusty clump (size $<$2\,pc) exhibits 
a strong 2175$\Angstrom$ bump.
They speculated that large, submicron-sized 
dust in M\,31 may have been destroyed 
in the harsh environments of the bulges 
by supernova explosions or past activities 
of the central super-massive black hole, 
resulting in the observed steepened extinction curve. 




\begin{figure}
\centering
\includegraphics[angle=0,width=4in]{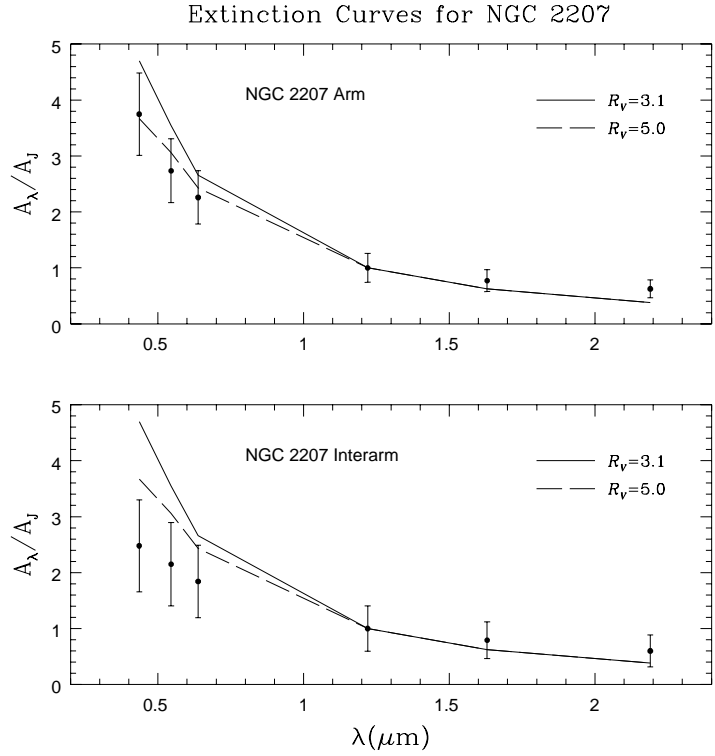}
\caption{\footnotesize
         Extinction curves in the $BVRJHK$ bands
         (normalized to the $J$-band extinction)
         for the dust in the spiral arm (top) 
         and interarm (bottom) regions of 
         NGC\,2207, a spiral galaxy partially 
         occulting the background spiral IC\,2163
         (Berlind et al.\ 1997).
         The solid and dashed lines plot
         the MW extinction curves for 
         $\RV=3.1$ and $\RV=5.0$, respectively.
         Note that the extinction curves in 
         both the spiral arm and interarm regions 
         of NGC\,2207 are flatter (``grayer'') 
         than the MW average of $\RV=3.1$,
         while the interarm extinction curve 
         is even ``grayer'' than 
         that of the spiral arm. 
         }
\end{figure}

\section{Dust in Galaxies beyond the Local Group}
The CCM extinction relation is applicable to 
a wide range of interstellar dust
environments in the MW, 
including lines of sight 
through diffuse dust and dark cloud dust,
as well as dust associated with star formation. 
However, the CCM relation does not appear to 
apply beyond the MW even in other Local Group 
galaxies such as the Magellanic Clouds and M\,31 
(see Clayton 2004).
Unfortunately, high signal-to-noise extinction
curves are still available for only three galaxies, 
the MW, LMC, and SMC.
Direct measurements of extragalactic UV extinction 
using individual reddened stellar sightlines 
are still limited to the LMC, SMC 
and a few sightlines in M\,31 (see \S5). 
As summarized by Draine (2003),
there are several different approaches 
to determining the extinction curve for 
dust in galaxies where individual stars 
cannot be resolved.


%
%

%
White \& Keel (1992) proposed that 
the dust extinction curve of 
an external galaxy
can be determined directly 
if it lies in front of another galaxy.  
The ideal case consists of a face-on 
symmetric spiral galaxy half overlapping 
with a symmetric, background spiral 
or elliptical galaxy.  
The galaxies need to be symmetric 
so that their non-overlapping parts 
may be used to estimate
the surface brightness 
of their overlapping parts.
This method has been applied to 
a number of overlapping galaxy pairs
(White \& Keel 1992; 
White et al.\ 1996, 2000;
Keel \& White 2001a,b).  

Using this technique, 
Berlind et al.\ (1997) measured
the extinction curves
in the $BVRJHK$ visual to near-IR wavelength range
for the dust
in the spiral arm and interarm regions 
of NGC\,2207 which overlaps IC\,2163. 
This pair consists of two symmetric, 
almost face-on interacting spiral galaxies 
which are partially overlapping. 
These galaxies have been observed 
and modeled in detail by D.~Elmegreen et al.\ (1995)
and B.~Elmegreen et al.\ (1995).

Berlind et al.\ (1997) found that the dust 
in NGC\,2207 is mainly concentrated in its 
spiral arms, leaving its interarm regions 
mostly transparent. More interestingly, 
they found that the extinction curve in 
the spiral arm region of NGC\,2207 is flat,
resembling the MW extinction curve for 
$\RV\approx5.0$, 
while the interarm dust appears 
to be even ``grayer'' (see Figure~10).
They proposed that an unresolved patchy dust 
distribution in NGC\,2207 could be capable of 
producing the observed extinction curves;
and the arm-interarm difference in the observed extinction 
could also be explained if the interarm region 
has a higher degree of dust patchiness 
(i.e., a larger density ratio between 
the high-density and low-density phases) 
than the arm region.


Finally, we note that 
there are some distant extragalactic point sources 
that can be used to construct extinction curves 
if their intrinsic SEDs are well characterized. 
These include QSOs and gamma-ray bursts (GRBs). 

By comparing the composite spectra 
of unreddened and reddened quasars
(Vanden Berk et al.\ 2001),
one can derive the extinction
curve for quasar sightlines.
Richards et al.\ (2003) 
and Hopkins et al.\ (2004)
found that the reddening toward thousands 
of SDSS quasars is dominated by SMC-type dust,
but some studies 
argue for ``gray'' dust
(see Li 2007).

Distant quasars have also been used 
as a background source to determine 
the dust extinction of 
damped Ly$\alpha$ absorption systems
(e.g., see Wang et al.\ 2012)
and intervening (e.g., Mg II) 
absorption systems 
(e.g., see Wang et al.\ 2004,
Zhou et al.\ 2010, 
Jiang et al.\ 2010a,b, 2011).

Gravitationally-lensed quasars 
with multiple images can also be 
used to determine the extinction 
curves of distant galaxies
(e.g., see Motta et al.\ 2002,
but also see McGough et al.\ 2005). 


GRBs, owing to their intense luminosity
(emitting up to $\simali$$10^{53}$\,erg), 
allow their detection up to very high redshifts. 
Particularly, 
the featureless, power law-like spectral shapes of 
their afterglows, make GRBs an excellent probe of 
the dust extinction for the GRB host galaxies
(e.g., see Li et al.\ 2008, Liang \& Li 2009, 2010).

We note that the 2175$\Angstrom$ extinction bump 
has been detected in both nearby and distant galaxies 
(see Xiang, Li, \& Zhong 2011 and references therein).








\vspace{3mm}
{\noindent {\bf Acknowledgements}~ We thank
R.A.~Benjamin,
D.L.~Block (DLB),
G.C.~Clayton,
B.G.~Elmegreen (BGE),
B.T.~Draine,
G.G.~Fazio,
K.C.~Freeman,
M.L.~Norman,
M. Rubio,
and A.N.~Witt for helpful discussions.
We are supported in part by NSFC\,11373015 and 11173007,
NSF AST-1109039, NASA NNX13AE63G,
and the University of Missouri Research Board.
It is a great pleasure
to acknowledge the important role that
DLB and BGE have played in pushing astrophysics
to the new front.
One of us (AL) would like to thank
the SOC for inviting him to attend this
fantastic, memorable conference in Seychelles.




\end{document}